\documentclass{PoS}

\usepackage{verbatim}

\title{New approach to the Dirac spectral density in lattice gauge theory applications}

\ShortTitle{New approach to the Dirac spectral density with applications}

\author{Zoltan Fodor\\
        University of Wuppertal, Department of Physics, Wuppertal D-42097, Germany\\
        Juelich Supercomputing Center, Forschungszentrum Juelich, Juelich D-52425, Germany\\
        \email{fodor@bodri.elte.hu}}

\author{Kieran Holland\footnote{Presenter,  with poster contributors Julius Kuti and Chik Him Wong.}\\	
        University of the Pacific, 3601 Pacific Ave, Stockton CA 95211, USA\\
        Albert Einstein Center for Fundamental Physics, Bern University, Bern, Switzerland\\
        \email{kholland@pacific.edu}}

\author{{Julius Kuti}\\
        University of California, San Diego, 9500 Gilman Drive, La Jolla, CA 92093, USA\\
        \email{jkuti@ucsd.edu}}

\author{Santanu Mondal\\
          Eotvos University, MTA-ELTE Lendulet Lattice Gauge Theory Group, Budapest, Hungary\\
        \email{santanu@bodri.elte.hu}}

\author{Daniel Nogradi\\
           Eotvos University, MTA-ELTE Lendulet Lattice Gauge Theory Group, Budapest, Hungary\\
        \email{nogradi@bodri.elte.hu}}

\author{ {Chik Him Wong}\\
        University of Wuppertal, Department of Physics, Wuppertal D-42097, Germany\\
        \email{cwong@uni-wuppertal.de}}

\abstract{We report tests and results
 ~from a new approach to the spectral density 
and the mode number distribution of the
Dirac operator in lattice gauge theories. The algorithm
generates the spectral density of the lattice Dirac operator as a continuous function 
over all scales of the complete eigenvalue spectrum. 
This is distinct from an 
earlier method where the integrated spectral density (mode number) was calculated efficiently 
for some preselected fixed range of the integration.
The new algorithm allows global studies like the chiral condensate from the Dirac spectrum at any scale 
including the cutoff-dependent IR and UV range of the spectrum.
Physics applications include the scale-dependent mass anomalous dimension, spectral representation 
of composite fermion operators, and the crossover transition from the $\epsilon$-regime of Random Matrix Theory 
to  the p-regime in chiral perturbation theory. 
We present thorough tests of the algorithm in the 2-flavor sextet SU(3) gauge theory that we continue 
to pursue for its potential as a minimal realization of the composite Higgs scenario.}

\FullConference{The 33rd International Symposium on Lattice Field Theory\\
		 14 -18 July  2015\\
		 Kobe International Conference Center, Kobe, Japan}

\begin{document}

\section{Introduction and brief history}

We introduce a new approach to the Dirac spectral density and mode number distribution 
in lattice gauge theories. The algorithm effectively generates the spectral 
density of the lattice Dirac operator as a continuous function over the entire range of the eigenvalue spectrum
in large lattice volumes. 
This is distinct from an 
earlier method~\cite{Giusti:2008vb} where the integrated spectral density (mode number) was calculated efficiently 
for some preselected fixed range of the integration  and averaged over gauge configurations.
Motivated by~\cite{Giusti:2008vb} we set the goal to calculate efficiently the spectral density 
over the entire Dirac spectrum which can be integrated over any range to generate the mode number distribution
on arbitrary scales in a single application to the gauge configuration before the average is taken over the gauge ensemble.
Just like the method introduced 
in~\cite{Giusti:2008vb} for gauge theory applications, our method is also rooted in known applications of the
Chebyshev expansion from approximation theory when combined with stochastic evaluation of operator traces 
in large vector spaces. 

We have been developing and testing the reported lattice gauge theory algorithm over the last few years with 
results appearing in our earlier publications including~\cite{Fodor:2014zca,Fodor:2015vwa}.  Our implementation of the algorithm 
itself was only presented for the first time in~\cite{Fodor:sakata} with similar material to the one presented at this conference. 
We hope to motivate new work by our successful and thorough tests of this new lattice gauge theory application, implemented 
with staggered lattice fermions in our case.
As an example, for extension to other type of lattice fermions, interesting new results were presented 
at this conference on the chiral condensate of the Dirac operator with domain wall lattice fermions using the same approach~\cite{Cossu:2016yzp}. 

Based on the poster we presented at this conference, we demonstrate the effectiveness of the 
new algorithm. The plots in this short report are  updated from the poster
for better illustration of the algorithm and for pedagogical purposes. Section 2 is a brief summary
of the method. In Section 3 we present some tests and implementations as applied to the spectral density of the 
chiral condensate and the mode number distribution. Section 4 is a brief illustration of physics applications where 
we present our first tests of the GMOR relation and the scale dependent mass anomalous dimension of 
the chiral condensate in the 2-flavor sextet SU(3) gauge theory that we continue to pursue
for its viability as a minimal realization of the composite Higgs scenario. 
The method we present projects interesting new applications for future studies. 

\section{Resolution of spectral lines from a stochastic Chebyshev expansion}

In~\cite{Giusti:2008vb} the projector operator was used for the determination of the mode number
of the Dirac spectrum. A rational approximation to the projector operator was estimated by a Chebyshev polynomial 
expansion. Here we will approximate the $\delta$-function  of the Dirac operator, 
with its Chebyshev-Jackson polynomial approximation
which will determine the moments of the spectral density to a preset high order. 
Similar methods have been found and referenced 
for the curious reader in~\cite{Cossu:2016yzp} from recent history of approximation theory. 
Adding to the historical perspective, we note some very early work applying
Chebyshev approximation to moments of spectral densities for Hamiltonian spectra~\cite{Wheeler:1974},
curiously with co-authors from the same institutional affiliation as one of us, but from an earlier era.
Our method can be viewed as a similar approximation but has broader scope and general applicability to a large class
of Hamiltonian problems with stochastic implementation from the modern computer era.

\begin{figure}
\begin{center}
	\begin{tabular}{cc}
		\includegraphics[width=0.48\textwidth]{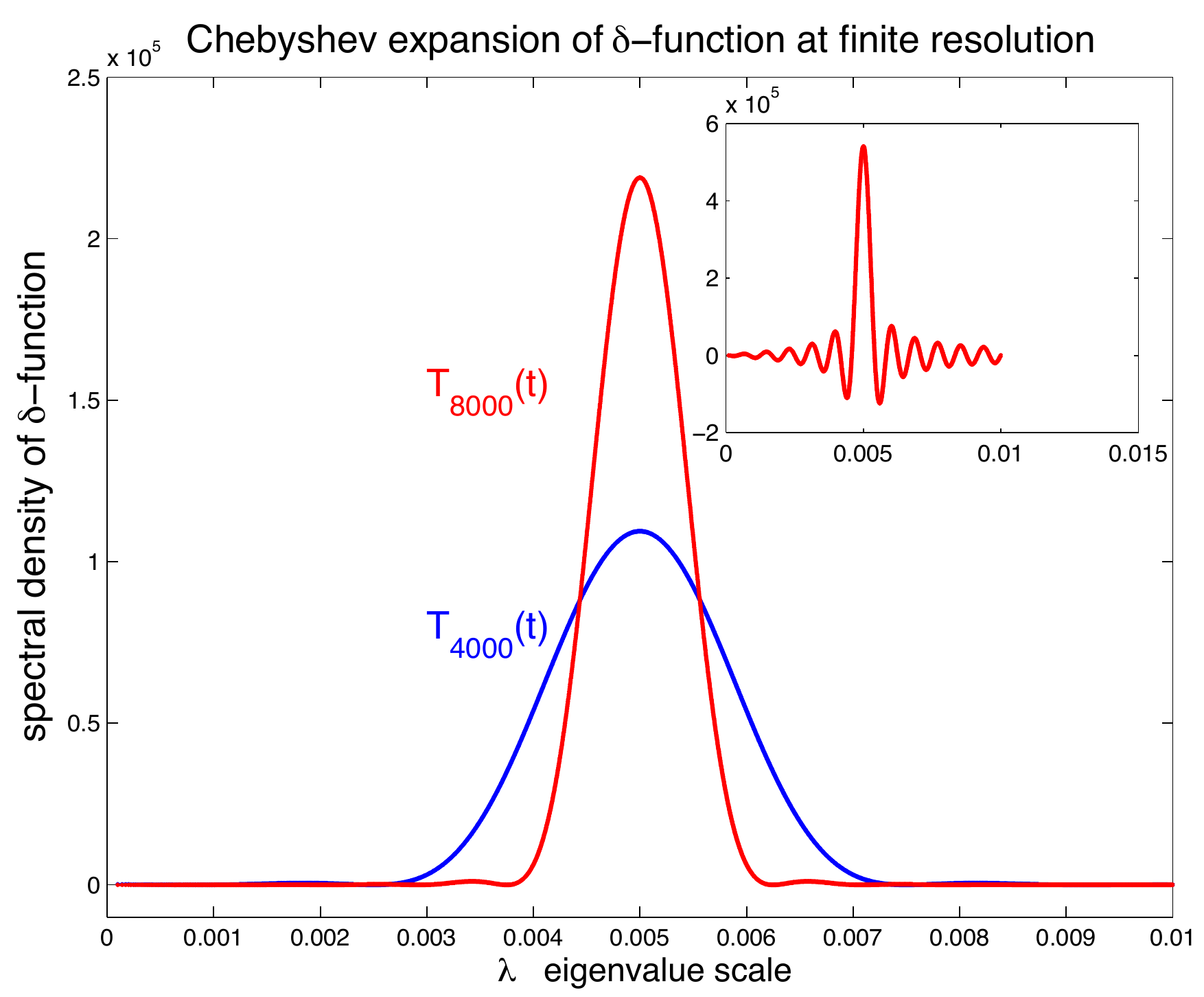}&
		\includegraphics[width=0.48\textwidth]{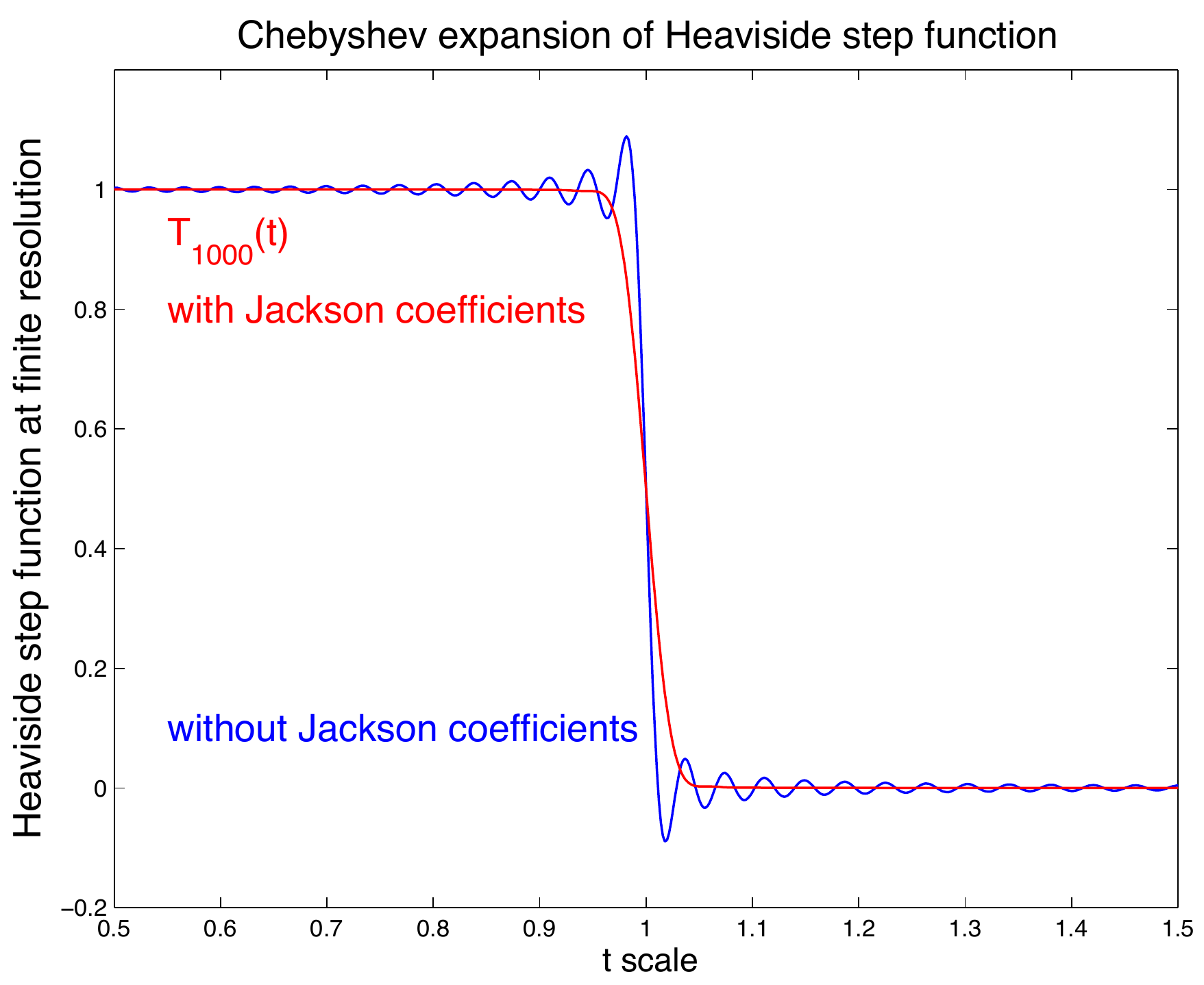}
	\end{tabular}
\end{center}
\caption{\footnotesize The left panel shows the sharp $\delta$-function spectral line in finite resolution with $1/N$ scaling and 
Gibbs oscillations (red insert) damped with Jackson coefficients in the main part of the left panel (red). 
The right panel shows the finite resolution of the Heaviside step
function.}
\label{fig:heaviside}
\end{figure}

We will describe the new algorithm for the spectrum of the staggered Dirac operator in a finite lattice volume, 
but the generalization to the spectrum of a large class of other operators
under some general spectral conditions is straightforward.
For a finite lattice four-volume $V$, with periodic or antiperiodic boundary conditions for fermions, the euclidean 
Dirac operator $D$ on any given gauge field configuration has purely imaginary 
eigenvalues $i\lambda_1,i\lambda_2, ...,$  with the associated average spectral density $\rho(\lambda,m)$,
\begin{equation}
\rho(\lambda,m) = \frac{1}{N_{eig}} \sum_{i=1}^{N_{eig}} \langle\delta(\lambda - \lambda_i) \rangle.
\label{eq:1}
\end{equation}
In Eq.~(\ref{eq:1})  the sum over $N_{eig}$ individual $\lambda_i$ eigenvalues is averaged over gauge configurations
which depend on the bare fermion mass $m$. 
The spectral density is a renormalizable quantity in gauge theories and the entire function in $\lambda$ 
can be computed at fixed $m$ on the lattice using the Chebyshev-Jackson expansion we will introduce. 
It is convenient to consider the mode number $\nu(\Lambda,m)$ of the positive definite 
hermitian operator $D^{\dagger}D+m^2$ in the integrated form of the spectral density as given by Eq.~(\ref{eq:2}),
\begin{equation}
\nu(M,m) = V \int_{-\Lambda}^{\Lambda} d\lambda \rho(\lambda,m), \hspace{5mm} \Lambda = \sqrt{M^2 - m^2}.
\label{eq:2}
\end{equation}
The  important role of the mode number distribution in the analysis of the chiral condensate
was emphasized  in~\cite{Giusti:2008vb} with a demonstration of its renormalization group invariance
$\nu_{\rm R}(M_{\rm R}, m_{\rm R}) = \nu(M,m)$.
After rescaling the spectrum of the Dirac operator $D$ and its equivalent  $D^{\dagger}D$ quadratic form, 
the spectral density $\rho(t)$ depends on 
scaled eigenvalues $\bar{\lambda}_i$  with
\begin{equation}
\rho(t) = \frac{1}{N_{eig}} \sum_{i=1}^{N_{eig}} \langle\delta(t - \bar{\lambda}_i) \rangle_{\it gauge~ensemble}
\end{equation}
where the variable $t$ is restricted to the [-1,1]  interval with rescaled eigenvalues $\bar{\lambda}_i$
in the [-1,1]  interval. 
The implicit dependence on the bare fermion mass $m$ often will not be shown for convenience. 
The density function $\rho(t)$ can be expanded into a series of $T_k(t)$ Chebyshev polynomials 
of the first kind,
\begin{equation}
\rho(t) = \frac{1}{\sqrt{1 - t^2}} \sum_{k=0}^\infty c_k T_k(t),
\end{equation}
with Chebyshev expansion coefficients
\begin{equation}
c_k = \left\{ 
\begin{array}{lll}
\frac{2}{\pi} \int_{-1}^{1} T_k(t) \rho(t) dt & & k = 0 \\ \\
\frac{1}{\pi} \int_{-1}^{1} T_k(t) \rho(t) dt & & k \ne 0 .
\end{array}
\right.
\label{eq:Cheb}
\end{equation}
In practical implementations $D$ will be replaced by the positive definite $D^{\dagger}D$ operator and the 
Chebyshev coefficients can be expressed in terms of $D^{\dagger}D$ eigenvalues,
\begin{equation}
c_k = \left\{ 
\begin{array}{lll}
\frac{2}{N_{eig} \pi} \sum_{i=1}^{N_{eig}} T_k(\bar{\lambda}_i^2)  & & k = 0 \\ \\
\frac{1}{N_{eig} \pi} \sum_{i=1}^{N_{eig}} T_k(\bar{\lambda}_i^2) & & k \ne 0.
\end{array}
\right.
\label{eq:3}
\end{equation}
Based on Eq.(\ref{eq:3}), the evaluation of the operator traces ${\rm Tr}(T_k(D^\dagger D))$ 
is needed to calculate the spectral density function. We use the well-known stochastic evaluation 
with Z(2) noise vectors $\xi$,
\begin{equation}
{\rm Tr}(T_k(D^\dagger D)) \approx \frac{1}{N_{\it noise}} \sum_{n=1}^{N_{\it noise}} 
\xi_n^T \cdot T_k(D^\dagger D) \cdot \xi_n .
\label{eq:noise}
\end{equation}
Recursion relations for Chebyshev polynomials of the operators,
\begin{eqnarray}
&& T_{k+1}(D^\dagger D)) \cdot \xi = 2 D^\dagger D \cdot T_k(D^\dagger D) 
\cdot \xi - T_{k-1}(D^\dagger D) \cdot \xi, \nonumber \\
&& \xi^{(n)} = T_n (D^\dagger D)) \cdot \xi^{(0)} \Leftrightarrow \xi^{(i+1)} = 2 D^\dagger D \cdot \xi^{(i)} - \xi^{(i-1)} ,
\end{eqnarray}
are used in averages over noise vectors $\xi$ in repeated recursions. The series has to be truncated at some finite 
order which will provide the Dirac $\delta$-function and the Heaviside step function $\Theta$ at finite resolution.
Figure~\ref{fig:heaviside} shows the spectral resolution of $\delta(\lambda)$ for a spectral line at $\lambda=0.005$
and $\Theta(t)$  at step $t=1$ for two different orders of the Chebyshev expansion. On the left panel, the resolution of 
the spectral function for a sharp $\delta$-function spectral line at $\lambda = 0.005$ is color coded. The width of the 
finite resolution 
is scaling with $1/N$ in the order $N$ of the Chebyshev expansion. In red color the resolution of the
spectral density at truncation $N=8000$ from the
Chebyshev coefficients of Eq.~(\ref{eq:Cheb}) is shown for the sharp $\delta$-function spectral line.
The truncation, as the insert of the left panel shows, introduces well-known Gibbs 
oscillations in the truncated spectral function at finite resolution. The Gibbs oscillations can be damped by the modified 
expansion 
\begin{equation}
\rho(t) = \frac{1}{\sqrt{1 - t^2}} \sum_{k=0}^\infty c_k T_k(t) \Rightarrow \frac{1}{\sqrt{1 - t^2}} \sum_{k=0}^\infty c_k g_k T_k(t) ,
\label{eq:Jackson}
\end{equation}
with Jackson coefficients $g_k$ which are well-known in approximation theory~\cite{Jackson:1930} for
damping the Gibbs oscillations with slight loss in the resolution. Blue color on the left panel of Figure~\ref{fig:heaviside}
shows the Chebyshev-Jackson expansion for the same spectral line at lower resolution consistent with $1/N$ scaling. 
Color coding on the right panel shows the resolution of the Heaviside step function comparing the Gibbs oscillation 
and its Jackson damping at the same Chebyshev order.

\section{Spectral density and Mode number}
\begin{figure}[htb!]
	\begin{center}
		\begin{tabular}{cc}
		 	\includegraphics[width=0.48\textwidth]{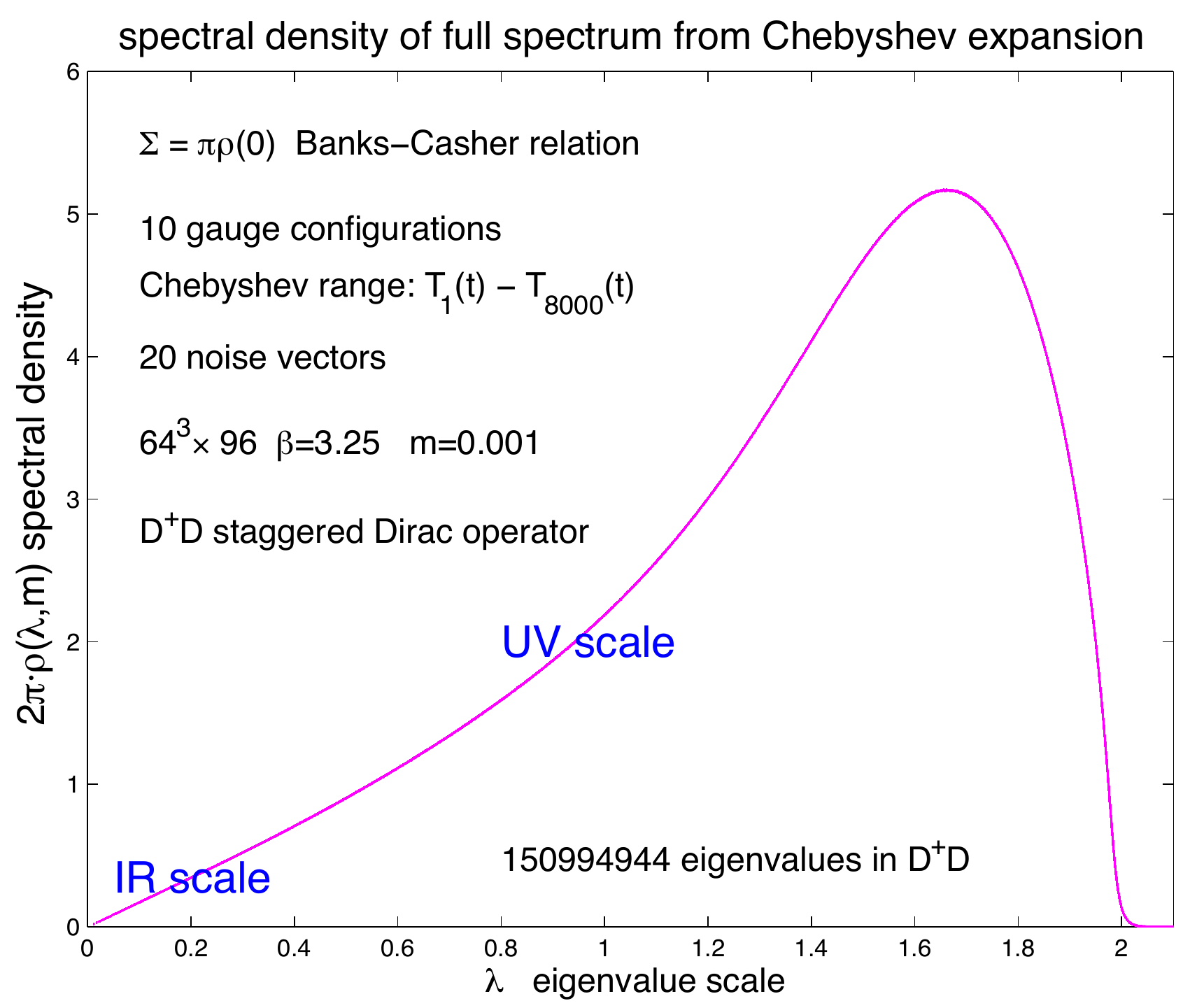}&
			\includegraphics[width=0.48\textwidth]{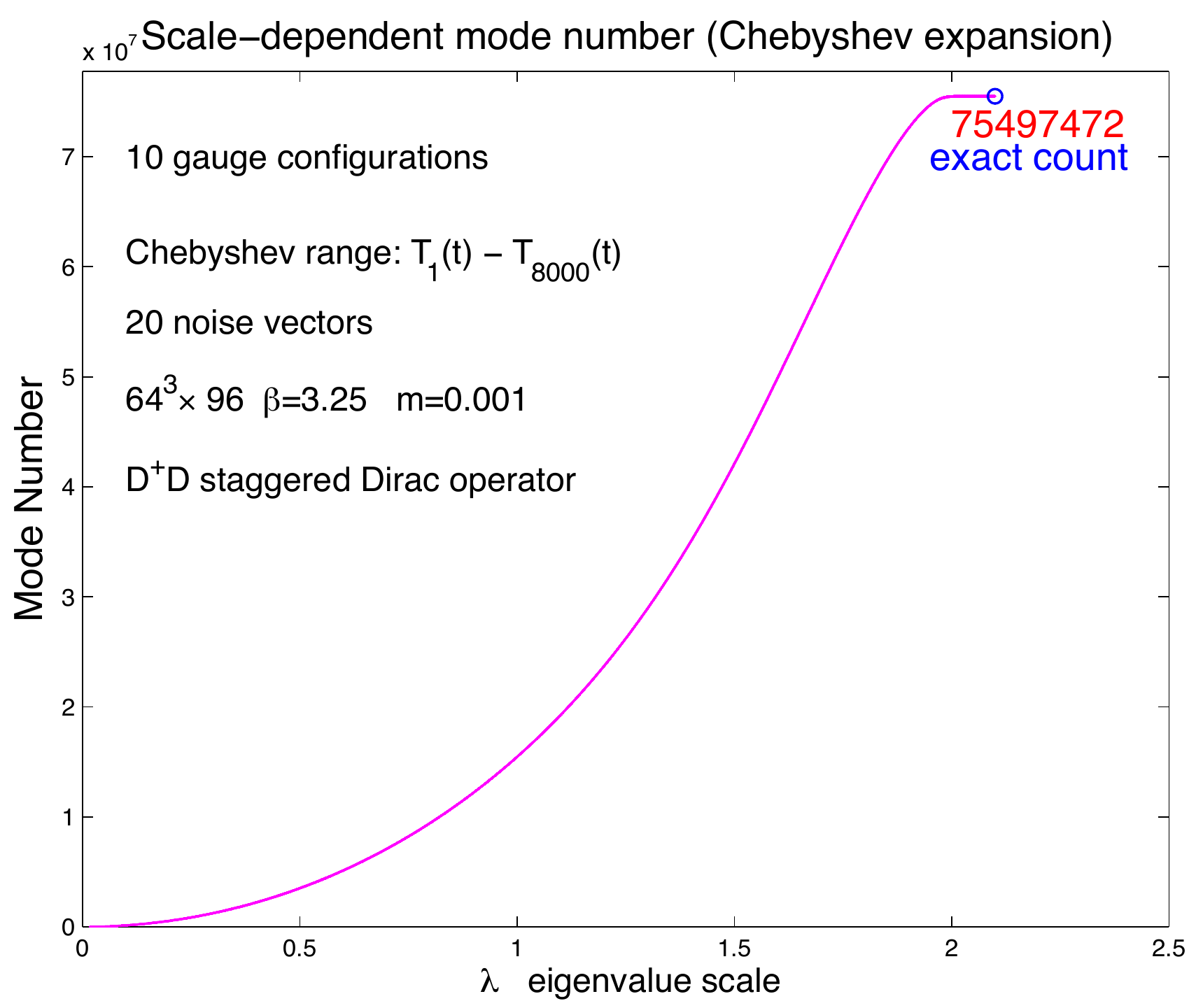}\\
        	\includegraphics[width=0.48\textwidth]{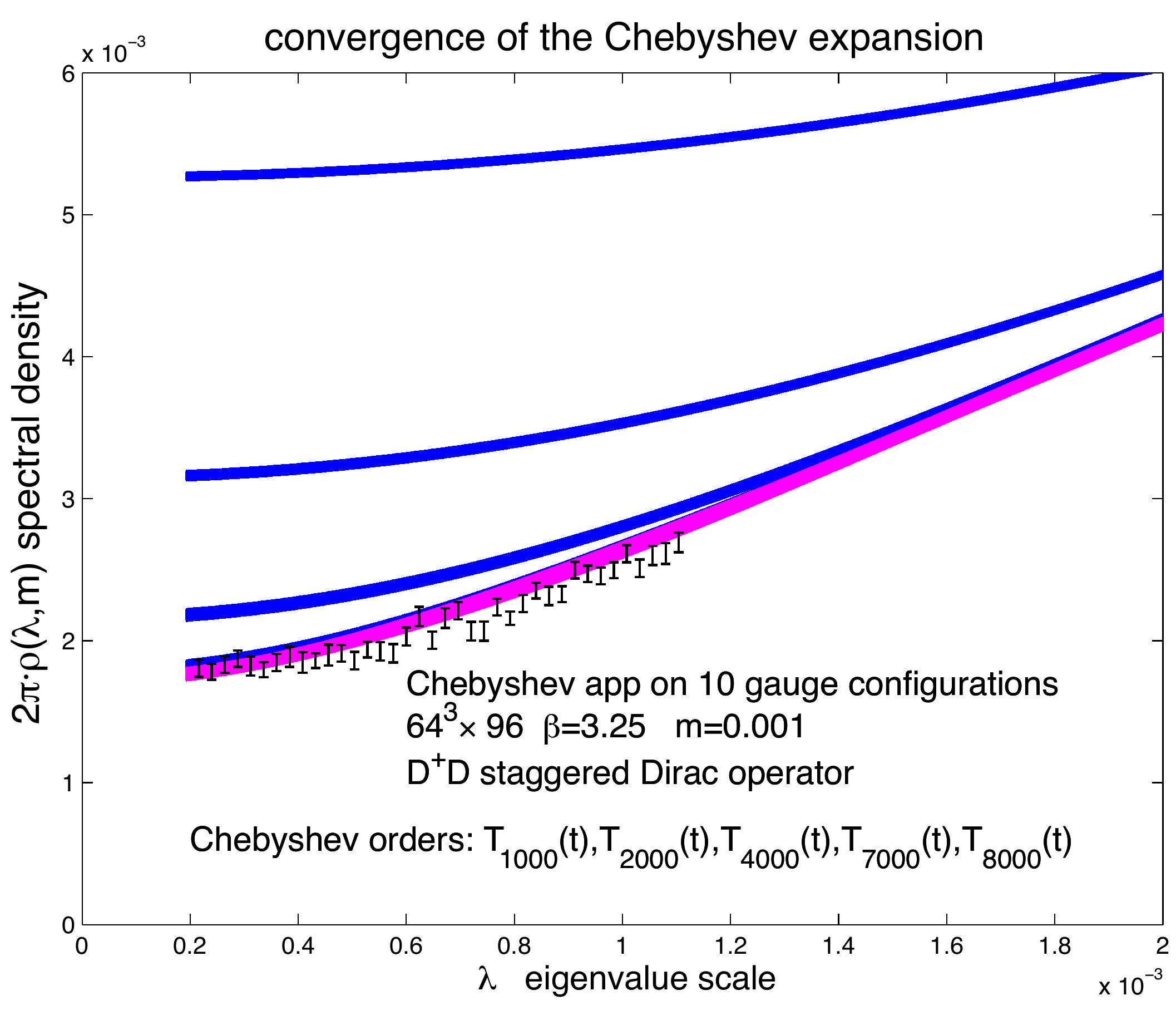}&
			\includegraphics[width=0.48\textwidth]{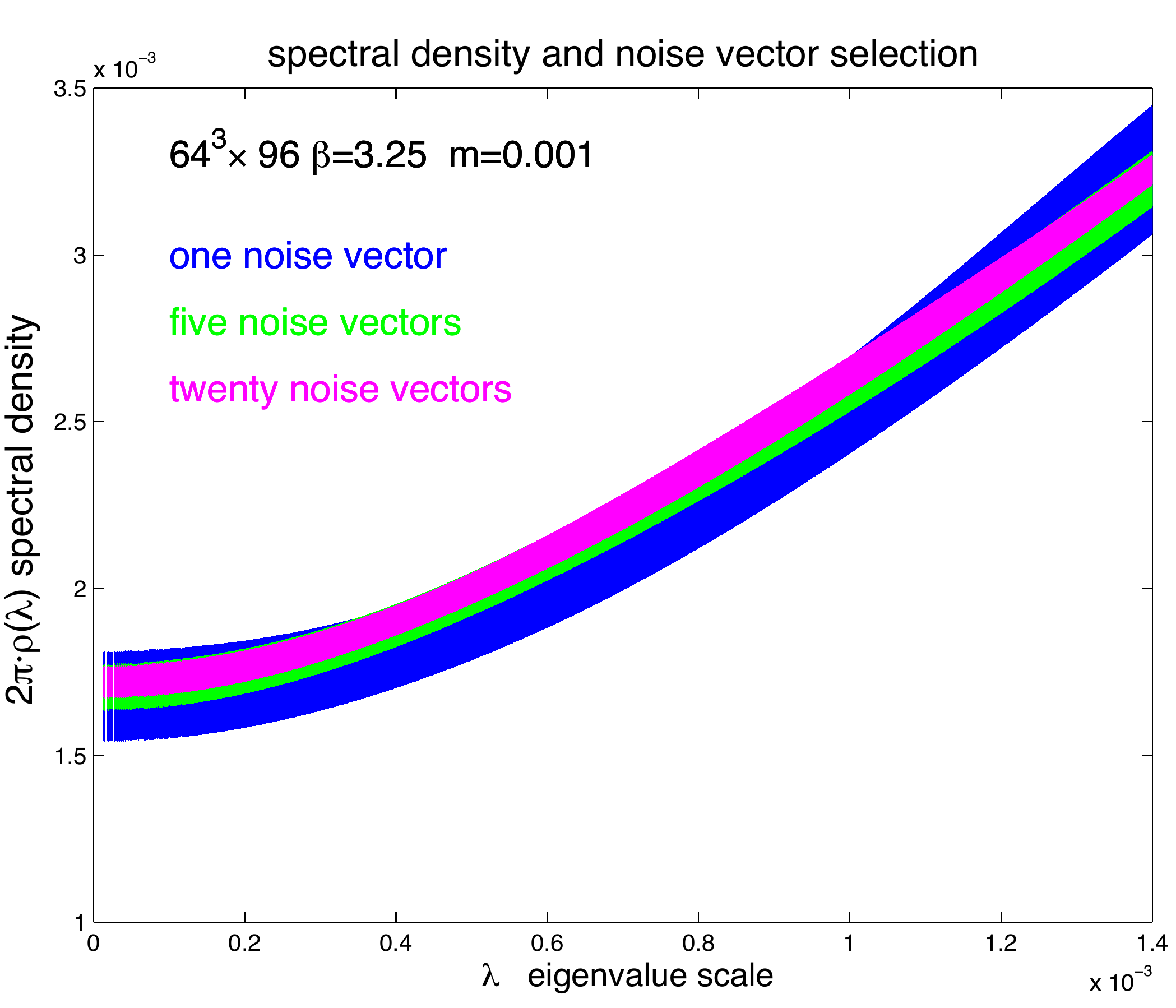}						
		\end{tabular}
	\end{center}
		\vskip -0.2in
	\caption{\footnotesize  Tests with a fermion  doublet in the two-index symmetric (sextet) representation 
of the SU(3) BSM color gauge group in the composite figure are discussed in the text. The black data points 
in the lower left panel come from direct diagonalization which is an important part of the testing procedure. The magenta line 
of the panel shows the result with N=8000 Chebyshev order in perfect match to the direct diagonalization data.}
\label{fig:spectrum}
\end{figure}

To illustrate the efficiency of the method,  
Figure~\ref{fig:spectrum} shows results from the calculation of the spectral density and the related mode number distribution
on all scales using the Chebyshev-Jackson expansion of Eq.~(\ref{eq:Jackson}).
The calculation targets here an
important BSM gauge theory with a fermion  doublet in the two-index symmetric (sextet) representation 
of the SU(3) BSM color gauge group as reviewed in talks at this conference~\cite{kuti:2016}.
The upper left panel of the figure shows the spectral density of the staggered $D^\dagger D$ Dirac operator 
using ten independent gauge configurations on the  lattice volume with size $64^3\times96$ at bare gauge 
coupling set by $\beta=3.25$ in the lattice action~\cite{kuti:2016} and the fermion mass set at $m=0.001$. 
Chebyshev polynomials up to order 8000 were used in the 
expansion with 20 noise vectors defined in Eq.~(\ref{eq:noise}).  The size of the Jackknife errors in the spectral density 
is not visible on the scale of the upper left panel of the plot. 
The upper right panel shows the mode number counting the eigenmodes from the integral of the spectral density.
It converges to the correct total count which is half of what is shown in the upper left panel from the 
implementation of $D^\dagger D$ on the staggered lattice counting only one of two degenerate 
eigenvalues of each pair in the spectrum.
The lower left panel magnifies the far infrared (IR) scale of the spectral density illustrating the convergence 
of the Chebyshev-Jackson expansion as the polynomial order is increased. This panel also illustrates that the convergence
rate is the slowest in the IR region and reached at polynomial order $N=8000$ with the magenta line which is 
practically identical to the lowest blue
line at $N=7000$ even in the lowest eigenvalue range. 
Bands in the magnified IR part of the plot show the visible Jackknife errors of the spectral density. 
All tests were made with 20 noise vectors. Data points from the direct diagonalization of $D^\dagger D$
are in excellent agreement with the expansion even in the low IR region. The lower right panel shows the test when 
the number of noise vectors is varied at fixed expansion order which is kept lower. The results show no bias as a function
of the noise vector number and with twenty (or fewer) noise vectors the variance is dominated by the fluctuations over the gauge ensemble. The increased size of the error band simply comes from the larger magnification.
%

%

\section{Physics applications: GMOR and mass anomalous dimension}

Excellent agreement between low eigenvalues 
from direct diagonalization of  $D^\dagger D$ and the full spectral density function from the Chebyshev-Jackson expansion,
as shown in Figure~\ref{fig:spectrum}, provides checks on the GMOR relation for the chiral condensate
as discussed in~\cite{Fodor:2014zca,Fodor:2015vwa}. 
\begin{figure}[h!]
\begin{center}
\includegraphics[width=.5\textwidth]{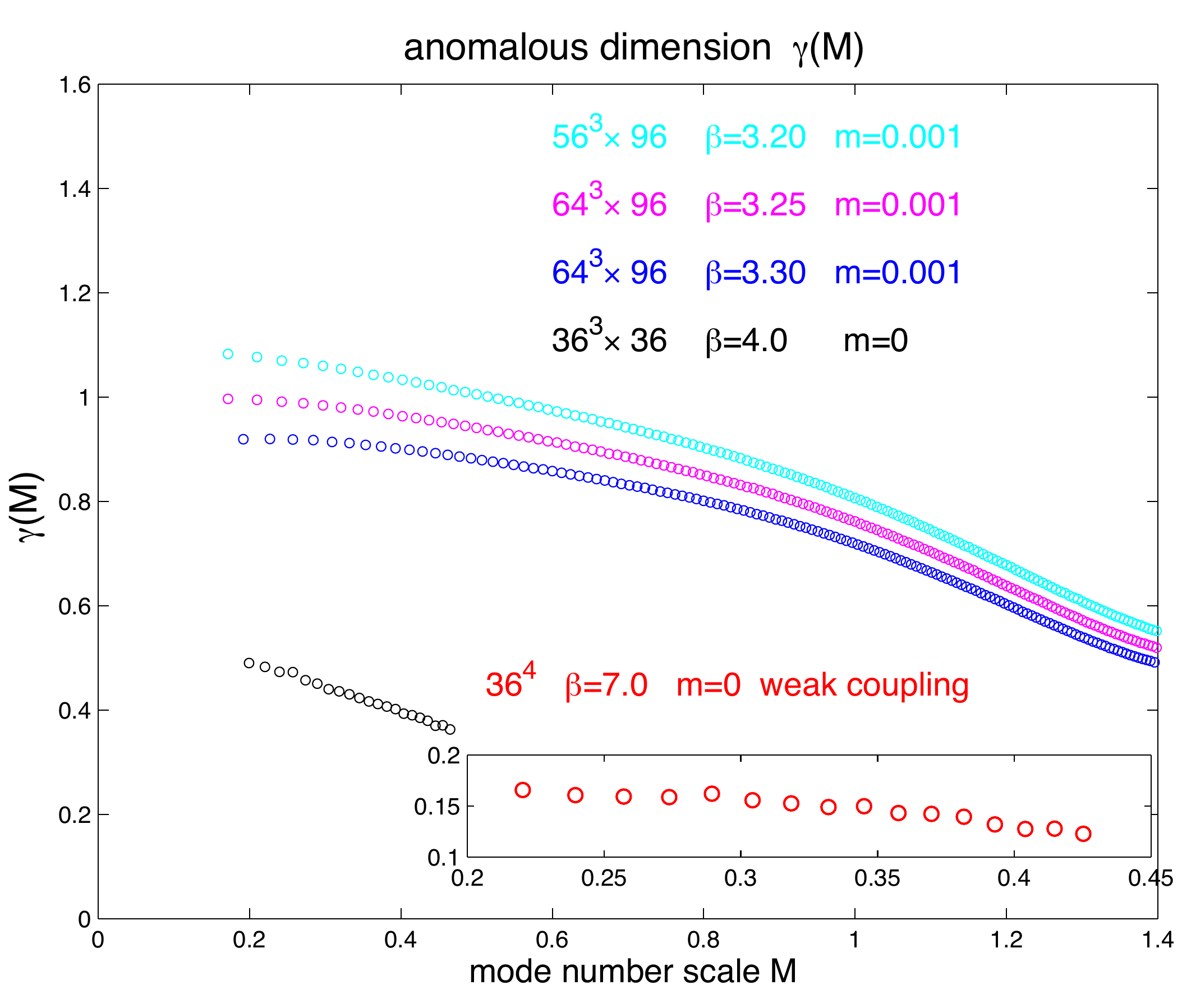}
\caption{\footnotesize Results for the mass anomalous dimension $\gamma(M)$,
as defined in the text,
are shown for the model with a fermion  doublet in the two-index symmetric (sextet) representation 
of the SU(3) BSM color gauge group. There are results at five different lattice spacings, one of them at exactly zero 
fermion mass. With the renormalization constants $Z_p$ and $Z_m$ determined in separate calculations, the 
continuum scale dependent mass anomalous dimension and its role in fermion mass generation is left for a future publication in preparation.}
\end{center}
\label{fig:anom_dim}
\vskip -0.2in
\end{figure}
Chiral perturbation theory of the effective chiral condensate $\Sigma_{\rm eff}$ , aided by the Chebyshev-Jackson
expansion was used in our tests of the
GMOR relation for added evidence of chiral symmetry breaking in the sextet BSM theory of the composite Higgs. 
The effective chiral condensate $\Sigma_{\rm eff}$ was analyzed based on 
Eq.~(\ref{eq:sigma}) as derived in~\cite{Osborn:1998qb},
%
%
 %
 \begin{equation}
 \frac{\Sigma_{\rm eff}}{\Sigma} = 1 + \frac{\Sigma}{32 \pi^3 N_F F^4} \left[ 2 N_F^2 |\Lambda| \arctan \frac{|\Lambda|}{m} - 4 \pi |\Lambda| - N_F^2 m \log \frac{\Lambda^2 + m^2}{\mu^2} - 4 m \log \frac{|\Lambda|}{\mu} \right] .
 \label{eq:sigma}
 \end{equation}

Additional probing of chiral symmetry breaking 
includes the 
study of the mass anomalous dimension $\gamma(M)$ as shown in Figure 3.
The anomalous dimension of the chiral condensate can be determined from access to the mass anomalous dimension
$\gamma(M)$ in the eigenmode function~\cite{DelDebbio:2010ze,Patella:2012da,Cheng:2013eu},
\begin{equation}
\nu_{\rm R}(M_{\rm R}, m_{\rm R}) = \nu(M,m) \approx {\rm const} \cdot M^{\frac{4}{1 + \gamma_m(M)}} 
\label{eq:gamma1}
\end{equation}
The results are reported in Figure 3 with details left for a future publication in preparation.

\acknowledgments
We acknowledge support by the DOE under grant DE-SC0009919,
by the NSF under grants 0970137 and 1318220, by the DOE ALCC award for the BG/Q Mira platform
of Argonne National Laboratory, by OTKA under the grant OTKA-NF-104034, and by the Deutsche
Forschungsgemeinschaft grant SFB-TR 55. Computational resources were provided by the Argonne Leadership Computing Facility under an ALCC award, by USQCD at Fermilab, by the University of Wuppertal, by The Juelich Supercomputing Center on Juqueen
and by the Institute for Theoretical Physics, Eotvos University. 
We are grateful to Szabolcs Borsanyi for his code development for the BG/Q platform. We are also 
grateful to Sandor Katz and Kalman Szabo for their code development for the CUDA platform \cite{Egri:2006zm}. KH wishes to thank the Institute for Theoretical Physics and the Albert Einstein Center for Fundamental Physics at the University of Bern for their support.

\end{document}